# Films of $Mn_{12}$-acetate deposited by low-energy laser ablation

J. Means[a], V. Meenakshi[a], R.V.A. Srivastava[a], W. Teizer[a], Al.A. Kolomenskii[a], H.A. Schuessler[a], H. Zhao[b] and K.R. Dunbar[b]

[a]Department of Physics, Texas A&M University, College Station, TX 77843-4242, USA
[b]Department of Chemistry, Texas A&M University, College Station, TX 77843-3012, USA

## Abstract

Thin films of the molecular magnet $Mn_{12}$-acetate, $[Mn_{12}O_{12}(CH_3COO)_{16}(H_2O)_4] \cdot 2CH_3COOH \cdot 4H_2O$, have been prepared using a laser ablation technique with a nitrogen laser at low laser energies of 0.8 and 2 mJ. Chemical and magnetic characterizations show that the $Mn_{12}$-acetate cores remain intact and the films show similar magnetic properties to those of the parent molecular starting material. In addition, the magnetic data exhibit a peak in the magnetization at 27 K indicating the creation of an additional magnetic phase not noted in previous studies of crystalline phases.

## 1. Introduction

A recent topic of great interest in materials science is that of single-molecule magnets, molecules which possess a large-spin ground state. These molecules have been fabricated and studied extensively [1]. One such molecular magnet is the $Mn_{12}$-acetate compound, first fabricated by Lis in 1980 [2]. The $Mn_{12}$-acetate molecule shows magnetic hysteresis at low temperatures due to slow paramagnetic relaxation [3],[4]. The integration of these magnetic molecules into magnetic storage or nano-devices will require the successful production of thin films of the material. However, the thermal instability of the $Mn_{12}$-acetate compound renders standard thermal deposition techniques impractical. Films have been made using Langmuir–Blodgett techniques [5], as well as a few other methods [6],[7],[8], but we propose a simpler method, in line with more recent thin-film deposition technology. Pulsed laser deposition (PLD) has been used extensively for the deposition of organic and inorganic thin films [9]. In the present work, this technique is applied to produce films of $Mn_{12}$-acetate which mostly maintain the magnetic and molecular properties of the crystalline starting material.

Previously, some of us have performed laser ablation of $Mn_{12}$-acetate samples with an excimer laser with energies ranging from 200 to 450 mJ [10], [11]. Films formed with the excimer laser were found to show partial fragmentation in the $Mn_{12}$-acetate molecules at high laser energies. Lowering the laser energy was shown to reduce the fragmentation. In the present study we used a nitrogen laser that ablated $Mn_{12}$-acetate at much lower pulse energy.

## 2. Experimental

In preparation for the laser ablation, the crystalline starting material was pressed into a pellet at a pressure of ∼19,000 Torr. The pellet was then mounted as an ablation target in a high vacuum chamber ($P \approx 5 \times 10^{-6}$ Torr), where substrates (typically glass, mica or $Si/SiO_2$) were mounted at a distance of about 5 cm from the target. Films were fabricated using two different settings: (1) a pulse frequency of 8 Hz at an energy of 0.8 mJ for 5.5 h, and (2) a pulse frequency of 20 Hz with an energy of 2 mJ for 3 h. The resulting films were then characterized using X-ray photoelectron spectroscopy (XPS) and superconducting quantum interference device (SQUID) magnetometry in order to test if the $Mn_{12}$-acetate molecules remained intact and therefore retained their inherent molecular magnetic properties.

## 3. Results and discussion

Chemical characterization of the films was carried out using XPS. Fig. 1, Fig. 2 and Fig. 3 show the XPS data for the Mn 2p, O 1s, and C 1s peaks, respectively, in the starting material and in films created using the two different sets of conditions mentioned above. All data were normalized to a value of 1 at the highest energy. In Fig. 1, we note that the peak positions for Mn have changed very little. The Mn 2p peaks at 652.0 and 640.2 eV correspond to the $2p_{1/2}$ and $2p_{3/2}$ core levels, respectively. These are in reasonable agreement with the values found by Kang et al. using photoemission spectroscopy [12]. The shift of the Mn 2p peak positions in the ablated samples relative to those in the starting material range from 0.5 to 0.7 eV. The experimental resolution of ∼0.5 eV indicates that these shifts could be due to a small change in the binding state of the manganese atoms within the film. These shifts are consistent with a change in the magnetic properties, as will be discussed below.

In Fig. 2, it should be noted that the starting material shows two oxygen peaks (one appears as a shoulder on the right). These correspond to the oxygen atoms which are bonded to the manganese ions in the core of the $Mn_{12}$-acetate molecule (528.1 eV) and to the oxygen atoms contained in acetate and water ligands (529.9 eV) surrounding the core. These values are also in reasonable agreement with those given by Kang et al. [12]. Small variations might be attributed to charging effects due to the insulating nature of the samples. There is a shift in the relative intensity of the two peaks in the ablated material versus that of the starting material. In the original $Mn_{12}$-acetate compound, the integrated area of the peak from the oxygen atoms in the core is significantly smaller than that of the other peak. In the ablated samples, the area of the higher-energy peak is roughly equal to or lower than that of the other peak. This may indicate a reduction in the amount of oxygen from the water and acetate ligands, as could be the case if some of the ligands are being lost in the ablation process.

Fig. 3 shows two carbon peaks in the XPS data for the crystalline starting material. The two peaks, with binding energies of 283.3 and 287.1 eV, arise from the two carbon atoms in the acetate ligand [12]. A reduction in intensity in both of these peaks in the ablated films gives further indication that some acetate ligands may be lost in the ablation process. All of the XPS data taken together provide evidence that the manganese cores remain intact while some of the acetate ligands are lost in the ablation. Since the molecular magnetic properties result primarily from the core, this conclusion would be supported by the retention of the major magnetic features of the material, as shown below. This is a central issue in this study. Furthermore, we note that the loss of acetate ligands could permit an enhanced interaction between neighboring molecules and thus be responsible for the formation of an additional magnetic phase with a higher blocking temperature, as may be observed below. It should be noted that, while the given values for the XPS peak positions may show small variations from those previously reported in the literature, comparison of the values found for the films to those of the starting material are in close agreement.

Magnetic characterization was performed using SQUID magnetometry. In the $Mn_{12}$-acetate starting material, a strong hysteresis was observed in the magnetization data at low temperature, indicating a locking of the magnetic state as reported earlier [3] (not shown). Also of note were steps in the hysteresis, which indicate the presence of quantum tunneling of the magnetization [13], [14]. Fig. 4 shows the magnetization, $M$, as a function of the applied field, $B$, at temperature T=1.8K for $Mn_{12}$-acetate which was ablated at 8 Hz and 0.8 mJ and subsequently scraped from the substrate. This sample exhibits hysteretic behavior. The reason for the narrowing of the hysteresis as compared to that of the starting material is the random orientation of the molecules in the ablated material [5], [13]. Consistent with prior results (Fig. 3 in Ref. [5]), we do not observe the customary resonant tunneling steps, which is also attributed to orientational disorder in the thin films. The crystalline structure of the starting material leads to more defined features. The mere presence of hysteresis, however, argues that there is superparamagnetic blocking of the magnetization below 3 K.

Fig. 5 shows the magnetic moment of the ablated material in a field of 50 Oe as a function of temperature for the same sample cooled at zero-field and cooled at a field of 50 Oe. In the zero-field cooled curve, we see a peak at T=~2.2K (see inset, Fig. 5) which corresponds to the onset of superparamagnetic behavior. This peak is normally observed at 3.0 K in crystalline $Mn_{12}$, but this temperature has been found to depend on whether the material is a single crystal or polycrystalline [2], [15]. The presence of some other minority magnetic species can also cause a shift in the blocking temperature [16]. In addition to the peak at 2.2 K, a maximum was observed at T=~27K. This extra peak may result from the formation of an additional magnetic phase during the ablation process. This new phase causes the material to undergo a second transition with a higher blocking temperature than that of the individual molecules. We propose three possible explanations for the additional phase: (1) it could arise from the formation of clusters of linked molecules; (2) it could be the result of a surface effect wherein interactions with the substrate alter the anisotropy of the molecules; or (3) it could be the result of the creation of a spin-glass state within the material. Any of these processes would be consistent with the removal of acetate ligands (as indicated by the XPS data) and thus a stronger interaction between the $Mn_{12}$-acetate core and its immediate environment.

## 4. Conclusion

This study has shown that nitrogen laser ablation is a viable technique for the creation of thin films of $Mn_{12}$-acetate. XPS data are consistent with the conclusion that, while some of the surrounding acetate ligands and water molecules are removed in the ablation process, the cores of the molecules remain intact. Magnetic measurements support this conclusion, with a blocking temperature in the 2–3 K range along with the characteristic hysteresis. There is also evidence of an additional magnetic phase that has not been observed in previous studies of crystalline $Mn_{12}$-acetate.

## Acknowledgements

We thank C. Berlinguette, V. Pokrovsky, N. Sinitsyn and W. Lackowski for helpful discussions. J. Means is supported by a Doctoral Research Fellowship from Sandia National Laboratories. The authors gratefully acknowledge the following support: (1) Robert A. Welch Foundation, (2) Texas Advanced Research Program (010366-0038-2001), (3) The US Department of Energy (DOE--DE-FG03-02ER45999), (4) National Science Foundation for support from Nanoscale Science and Engineering (NIRT) Grant (DMR-0103455) and for an equipment grant to purchase a SQUID magnetometer (NSF-9974899).

# Figures

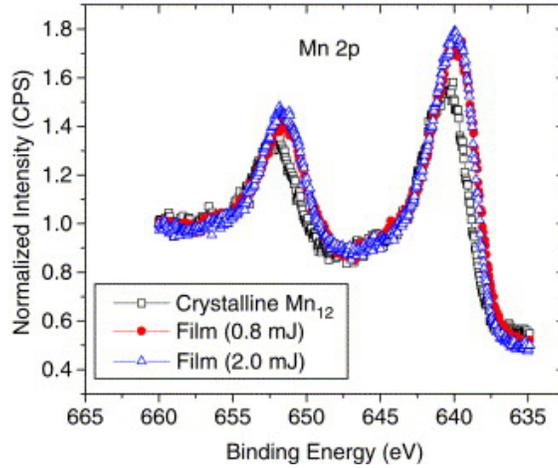

Fig. 1. XPS spectrum showing the binding energy peaks for the Mn 2p orbitals in the Mn$_{12}$-acetate starting material and two ablated films (one created at 8 Hz, 0.8 mJ, and the other at 20 Hz, 2.0 mJ), normalized to 1 at the highest binding energy.

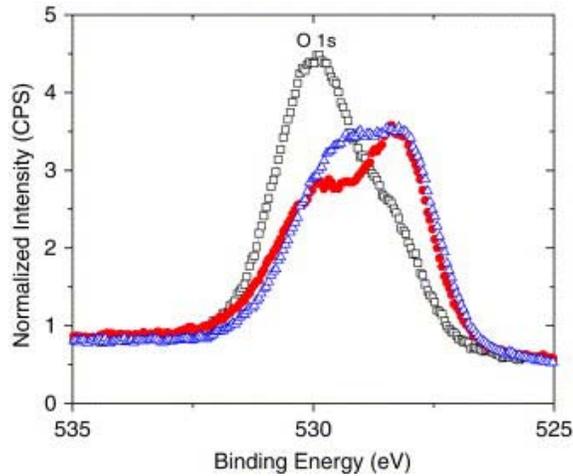

Fig. 2. XPS spectrum showing the binding energy peaks for the O 1s orbitals in the Mn$_{12}$-acetate starting material and two ablated films (one created at 8 Hz, 0.8 mJ, and the other at 20 Hz, 2.0 mJ), normalized to 1 at the highest binding energy. See Fig. 1 for key to symbols.

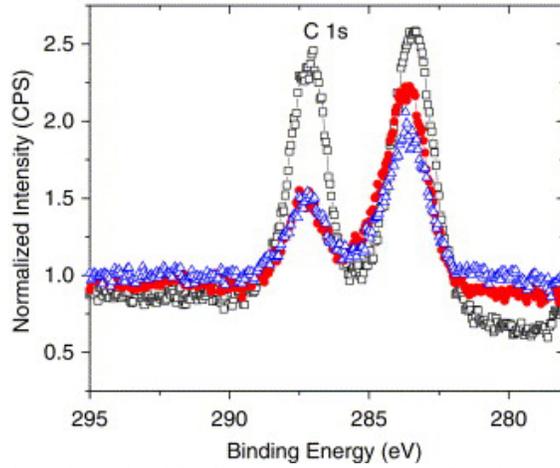

Fig. 3. XPS spectrum showing the binding energy peaks for the C 1s orbitals in the $Mn_{12}$-acetate starting material and two ablated films (one created at 8 Hz, 0.8 mJ, and the other at 20 Hz, 2.0 mJ), normalized to 1 at the highest binding energy. See Fig. 1 for key to symbols.

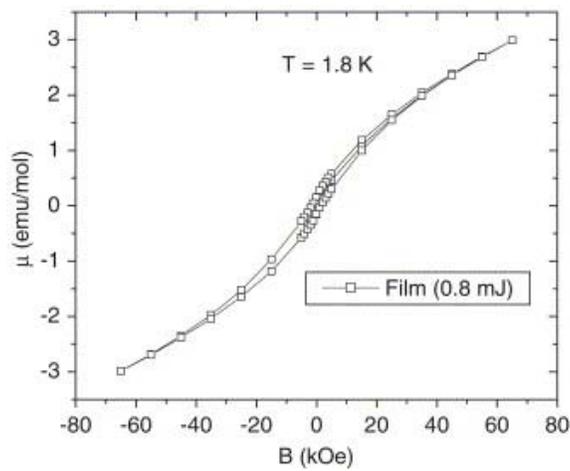

Fig. 4. Magnetization as a function of applied magnetic field for the ablated material at a temperature of 1.8 K, showing hysteresis.

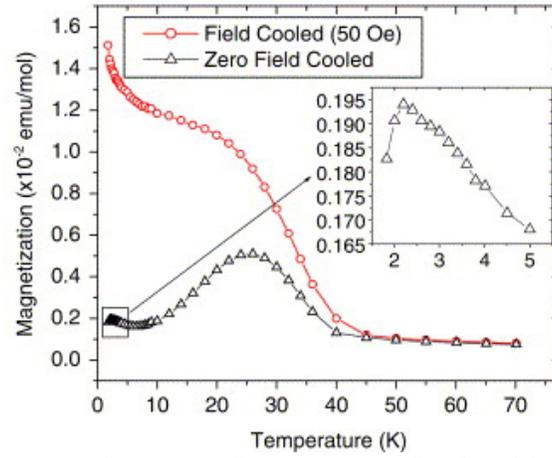
Fig. 5. Magnetic moment as a function of temperature for the ablated material in a field of 50 Oe with cooling in zero-field and in a field of 50 Oe. The insert shows the low-temperature region of the zero-field cooled data.